\documentclass[useAMS,usenatbib]{mn2e}
\usepackage[pdftex]{graphicx}

\newcommand{\Ha}{\mbox{H$\alpha$}}

\newcommand{\hd}{HD\,191612}

\newcommand{\beq}{\begin{equation}}
\newcommand{\eeq}{\end{equation}}

\newcommand{\fcl}{\mbox{$f_{\rm cl}$}}

\newcommand{\degree}{\mbox{\ensuremath{^\circ}}}

\newcommand{\vinf}{\mbox{$v_\infty$}}
\newcommand{\mdotun}{\mbox{$10^{-6}\,{\rm M_{\odot}/yr}$}}   

\newcommand{\rstar}{\mbox{$R_\star$}}
\newcommand{\rstun}{\mbox{$R_\odot$}}

\newcommand{\teff}{\mbox{$T_{\rm eff}$}}
\newcommand{\logg}{\mbox{$\log g$}}

\title[Dynamical Magnetospheres]{A dynamical magnetosphere model for
  periodic $\Ha$ emission from the slowly rotating magnetic O star
  \hd}

\author[J.O. Sundqvist et al.]{Jon O. Sundqvist$^{1}$\thanks{E-mail:
    jon@bartol.udel.edu}, Asif ud-Doula$^2$, Stanley
  P. Owocki$^{1}$, Richard H. D. Townsend$^{3}$\newauthor Ian D. Howarth$^4$, 
  Gregg A. Wade$^5$, and the MiMeS Collaboration\\ 
  $^1$University of Delaware, Bartol Research Institute, 
  Newark, Delaware 19716, USA\\ 
  $^2$Penn State Worthington Scranton, 120 Ridge View Drive, Dunmore, 
  PA 18512, USA\\
  $^3$University of
  Wisconsin, Department of Astronomy, Madison, WI 53706, USA\\
  $^4$University College London, Department of Physics and Astronomy,  
  Gower Place, London WC1E 6BT, United Kingdom\\
  $^5$Royal Military College of Canada, Department of Physics,  
  PO Box 17000 Kingston, Ontario K7K 7B4, Canada\\}

\begin{document}

\date{Accepted 2012-02-28. Received 2011-12-11}

\pagerange{\pageref{firstpage}--\pageref{lastpage}} \pubyear{2002}

\maketitle

\label{firstpage}

\begin{abstract}

The magnetic O-star \hd~exhibits strongly variable, cyclic Balmer line
emission on a 538-day period. We show here that its variable $\Ha$
emission can be well reproduced by the rotational phase variation of
synthetic spectra computed directly from full radiation
magneto-hydrodynamical simulations of a magnetically confined wind.
In slow rotators such as \hd, wind material on closed magnetic field
loops falls back to the star, but the transient suspension of material
within the loops leads to a statistically overdense, low velocity
region around the magnetic equator, causing the spectral
variations. We contrast such ``dynamical magnetospheres'' (DMs) with
the more steady-state ``centrifugal magnetospheres'' of stars with
rapid rotation, and discuss the prospects of using this DM paradigm to
explain periodic line emission from also other non-rapidly rotating
magnetic massive stars.

\end{abstract}


\section{Introduction}
\label{intro}

Shortly after \citet{Donati06} detected a strong magnetic field in the
Galactic Of?p star \hd, \citet{Howarth07} demonstrated that the
variable equivalent widths of its optical Balmer and He\,I lines
\citep[e.g.,][]{Walborn03} can be accurately phased according to a
538-day period, where in particular the outstanding $\Ha$ variation
shows strict periodicity. Since this period is unrelated to the much
longer orbital period $P_{\rm orb} = 1542 \, \rm d$ of \hd~and its
binary companion \citep{Howarth07}, rotational modulation of a
magnetically confined wind seems the most likely origin for the
variability, as already suggested by \citet{Donati06}. But in contrast
to centrifugally supported magnetosphere models, which have been
successfully applied to Balmer line variability in rapid rotators such
as the B star $\sigma$ Ori E \citep{Townsend05b}, it is not clear how
a very slow rotator such as \hd~can sustain a magnetosphere with
sufficient accumulation of wind plasma to explain the strong
  and periodic Balmer emission.

To reproduce the $\Ha$ variation of \hd, \citet{Howarth07} suggested
two geometrical toy models. One of these was indeed inspired by the
plasma distribution qualitatively expected from a magnetically
confined wind; it is a tilted, limb-darkened, geometrically thin disc,
where the sum of observer inclination $i$ and obliquity $\beta$ (the
angle between the rotation and magnetic axes) must be $i + \beta
\approx 100 \degree$ for the $\Ha$ modulation to be fit.

\citet{Wade11a} recently analysed Stokes V spectra of \hd. Assuming a
dipole oblique rotator, these authors derived $i+\beta = 95 \pm 10
\degree$, and by matching electron scattering modelling to the
observed photometric variability further obtained $i \ge 30
\degree$. A tentative reference geometry $i=30 \degree$ and $\beta= 67
\pm 5 \degree$ was then suggested from speculating that the orbital
and spin angular momenta of \hd~be aligned, and a surface dipole
(polar) field $B_{\rm d}=2450 \pm 400 \ \rm G$ derived.
  
This Letter examines to what extent full radiation
magneto-hydrodynamical (MHD) simulations of a magnetically confined
wind, along with detailed radiative transfer calculations, can
actually reproduce \hd's observed $\Ha$ variability, under the wind,
magnetic, and geometric constraints derived by \citet{Howarth07} and
\citet{Wade11a}.

\section{$\Ha$ in a spherically symmetric wind model}
\label{spherical}

\begin{figure}
\centering
\includegraphics[width=4.5cm, angle=90]{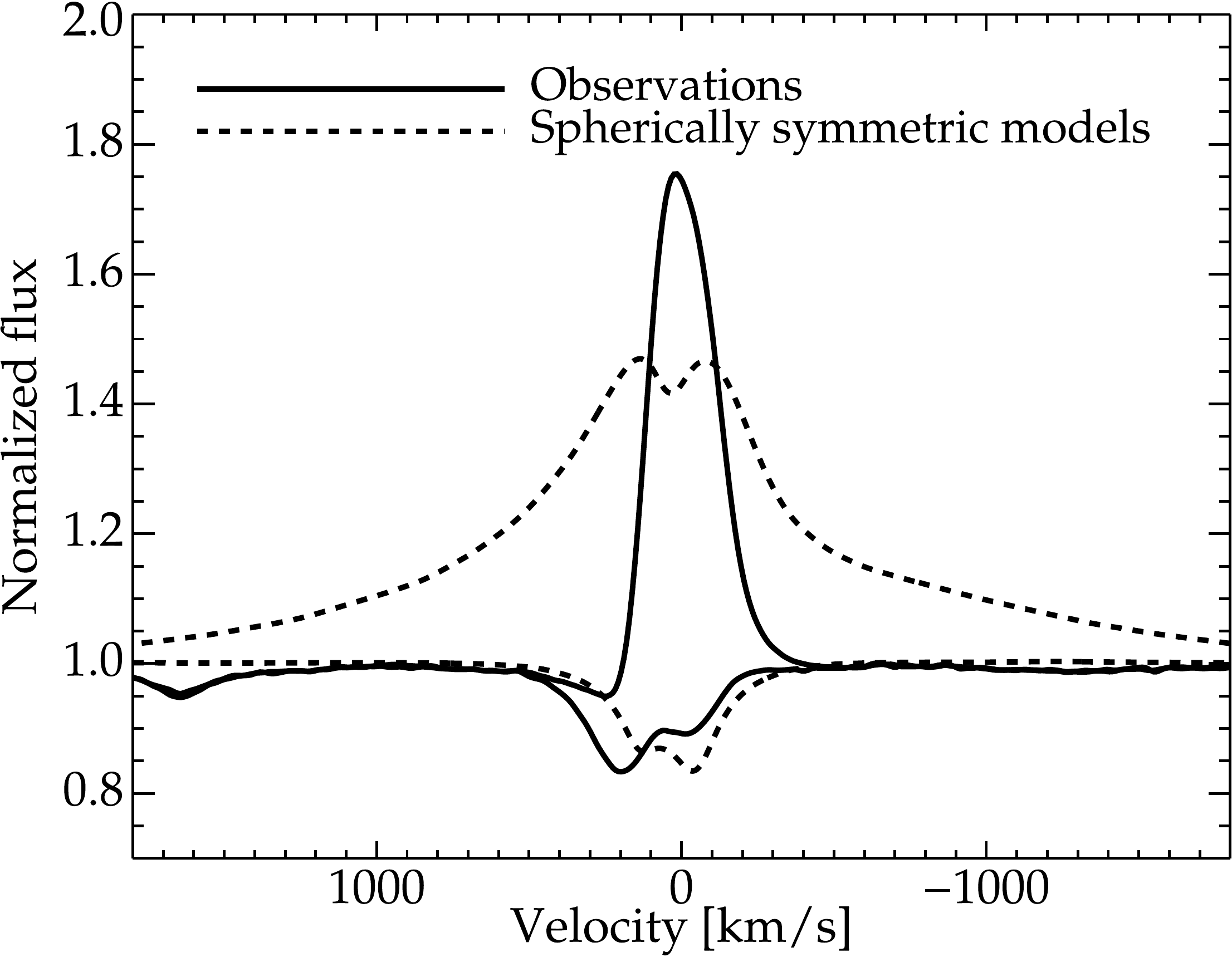}
\caption{Observed \citep{Howarth07} and synthetic $\Ha$ spectra during
  phases close to minimum and maximum. Synthetic {\sc fastwind}
  spectra are computed for two different mass-loss rates under the
  assumption of spherical symmetry (see Sect.~\ref{spherical}).}
\label{Fig:ha_sym}
\end{figure}

\begin{table}
	\centering
	\caption{Summary of stellar, wind, and magnetospheric parameters of \hd~\citep{Howarth07, Wade11a}.}
		\begin{tabular}{p{2.8cm}ll}
		\hline \hline Name & Parameter & Value \\ \hline
                Effective temperature
                & $\teff$ & $35\,000 \pm 1\,000 \, \rm K$ \\ Surface
                gravity & $\logg$ & $3.5 \pm 0.1$ \\ Stellar radius &
                $\rstar$ & 14.5\,$\rstun$ \\ Helium abundance &
                $n_{\rm He}/n_{\rm H}$ & 0.1
                \\ Terminal speed & $\vinf$ & 2700\,km/s \\ Mass-loss
                rate and clumping factor & $\dot{M} \sqrt{\fcl}$ &
                $1.6\,\times\,\mdotun$ \\ Surface polar magnetic field
                & $B_{\rm d}$ & $2450 \pm 400 \, \rm G$ 
                \\ Obliquity and observer inclination &
                $\beta + i$ & $95 \pm 10 \degree$ 
                \\
                \hline
		\end{tabular}
	\label{Tab:params}
\end{table}

To set the stage, we first compute synthetic $\Ha$ profiles for two
different mass-loss rates using the spherically symmetric, unified
(photosphere+wind) NLTE (=Non Local Thermodynamic Equilibrium) model
atmosphere code {\sc fastwind} \citep{Puls05}, taking stellar and wind
parameters from \citet{Howarth07}
(Table~\ref{Tab:params}). Fig.~\ref{Fig:ha_sym} confronts such models
with the observed $\Ha$ line profiles during minimum and maximum
phases.

The model-fit during minimum phase is acceptable. Indeed, it is
equivalent to that of \citet{Howarth07}, from which $\dot{M}
\sqrt{\fcl} = 1.6 \, \times \, \mdotun$ was derived for \hd. Here
$\fcl \equiv \frac{\langle \rho^2 \rangle}{\langle \rho \rangle^2} \ge
1$ is the wind clumping factor, with average mass density $\langle
\rho \rangle$. $\fcl$ enters the analysis because \Ha~is a
recombination based line in O stars, which means that its line
strength is greater in a small-scale structured (`clumped') wind than
in a smooth wind with the same mass-loss rate
\citep[e.g.,][]{Sundqvist11b}.

In an attempt to also match the maximum phase, we next calculated a
test-model with 5 times higher mass-loss rate. However,
Fig.~\ref{Fig:ha_sym} clearly shows that this synthetic $\Ha$ profile
is much broader than the observed one. This indicates that the
variable $\Ha$ emission does not stem from a variable global mass-loss
rate accompanied by a spherically symmetric velocity field such as the
$v = v_\infty (1-1/r)^\beta$ field assumed in {\sc fastwind}, here
with $\beta = 1$. Rather the variability might be caused by a confined
region of wind material with high density and low velocity; such
confinement may indeed stem from the strong magnetic field in
\hd~channeling its radiatively driven wind outflow to form a stellar
magnetosphere. We now describe our efforts to model this hypothesized
structure.

\section{Simulations}

\subsection{Modelling a dynamical magnetosphere}
\label{MHD}

Following the general procedure outlined by \citet{udDoula02}, we
compute a 2-D radiation MHD wind simulation of \hd, assuming a dipole
magnetic field. Hydrodynamical variables are specified on a standard,
right-handed spherical grid $(r,\theta,\phi)$, defined relative to a
Cartesian set $(x,y,z)$, where we assume symmetry in $\phi$. The
energy equation is treated as by \citet{Gagne05} and the radiation
line force is calculated within the Sobolev approximation using
standard CAK \citep{Castor75} theory. Since the rotation of \hd~is
extremely slow, the inferred period of 537.6 days \citep{Howarth07}
implies an equatorial rotation speed $v_{\rm rot} = 1.4 \, \rm km/s$,
we may neglect rotational effects on the dynamics (and thus use the
same simulation for any choice of obliquity $\beta$).

\begin{figure}

\centering
\begin{minipage}{6.8cm}
\resizebox{\hsize}{!}{\includegraphics[angle=90]{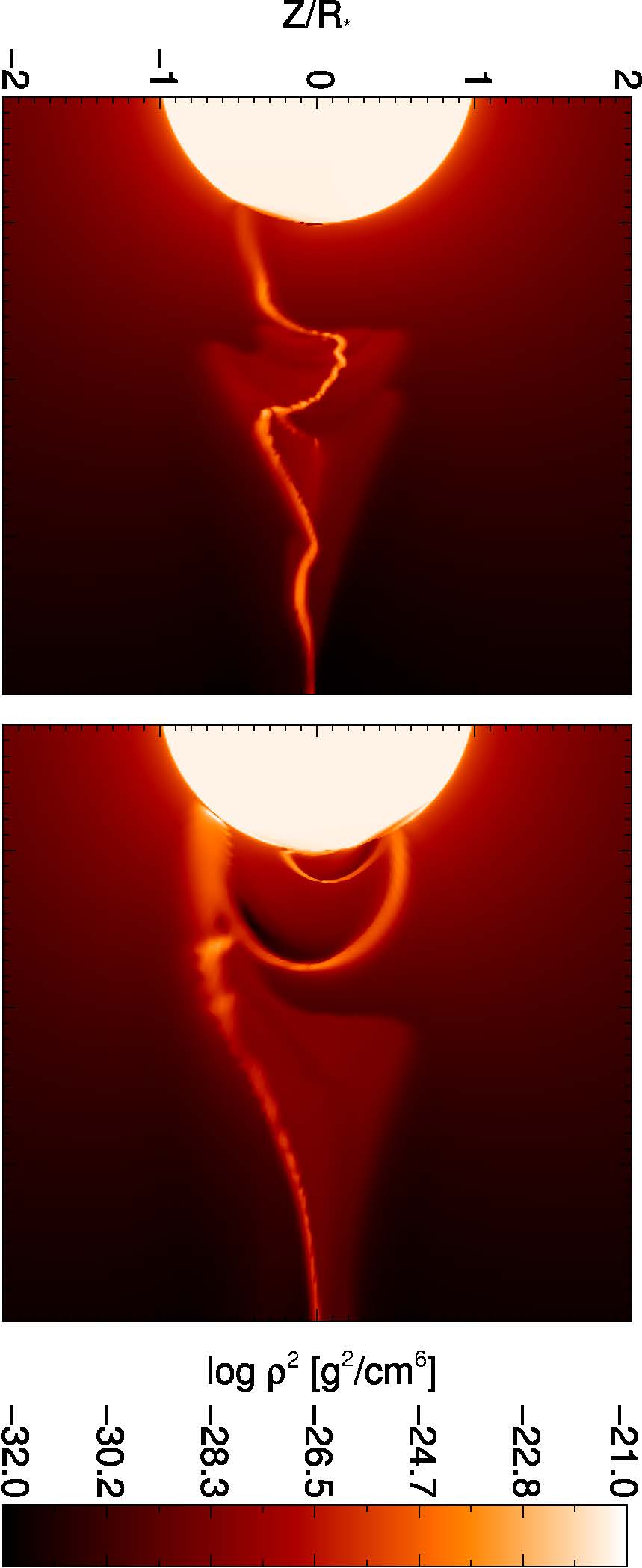}}
\end{minipage}

\centering
\begin{minipage}{6.8cm}
\resizebox{\hsize}{!}{\includegraphics[angle=90]{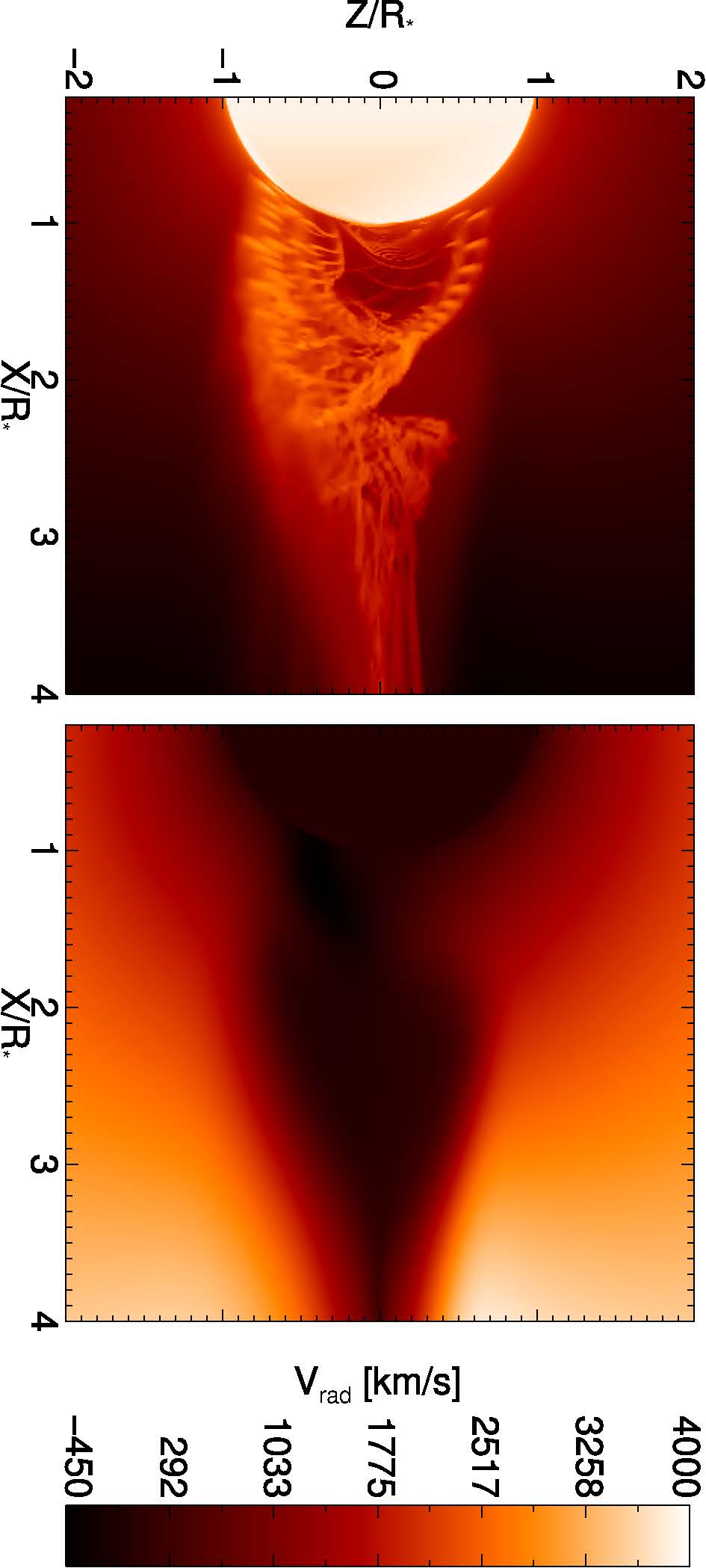}}
\end{minipage}
\caption{Contours of the density-squared for two different snapshots
  of the MHD wind simulation (upper panels), and of time-averaged
  density-squared (lower left) and radial velocity (lower right). Time
  averages are calculated from $>$ 100 snapshots taken well after the
  simulation's initial state. The Alfv\'{e}n radius is here located at
  $r \approx 2.7 R_\star$ whereas the Kepler co-rotation radius is at
  $r \approx 55 R_\star$, i.e. outside the range of the plots.}
\label{Fig:dens_sn}
\end{figure}

The effectiveness of the magnetic field in channeling the stellar wind
outflow may be characterized by the ratio of magnetic to wind kinetic
energy density,
\beq 
\eta \equiv \frac{B^2/8 \pi}{\rho v^2/2} = \eta_\star \frac{(r/R_\star)^{-4}}{v(r)/v_\infty},
\label{Eq:eta}
\eeq  
where the second equality defines the so-called `wind confinement
parameter' $\eta_\star \equiv B_\star^2R_\star^2/(\dot{M} v_\infty$)
\citep{udDoula02}, with $B_\star$ the dipole equatorial surface field
strength. If $\eta_\star > 1$, the dipole Alfv\'{e}n radius $R_{\rm A}
\approx \eta_\ast^{1/4} R_\ast$, at which the magnetic and wind energy
densities are equal, is located away from the stellar surface,
allowing then for some wind material to be channeled along closed
loops towards the magnetic equator (see Fig.~\ref{Fig:dens_sn}). But
the much steeper radial decline of the dipole magnetic energy density
($\sim 1/r^6$) than the wind kinetic energy density ($\sim 1/r^2$),
means that at large enough radii the wind will always force the field
lines to open up and essentially follow the radial wind flow.

The simulation here assumes strong confinement, $\eta_\star = 50$, in
accordance with the magnetic field strength recently derived by
\citet{Wade11a} and the wind parameters derived by \citet{Howarth07},
adopting $\fcl=1$. It is well established that the winds of hot,
massive O stars are indeed clumped \citep[see][for a recent
  review]{Sundqvist11b}. But a theoretical development of such
stochastic, small-scale inhomogeneities, as caused by the strong
instability inherent to line-driven winds \citep[e.g.,][]{Owocki88},
requires a non-Sobolev treatment of the radiation line force, and has
yet to be implemented within any MHD simulation. However, in terms of
the $\Ha$ modelling that is the focus of this paper, we are still
effectively modelling $\dot{M} \sqrt{\fcl}$ (see
Sect.~\ref{spherical}), but simply neglecting any \textit{dynamical}
effects such stochastic, small-scale structures might have upon the
large-scale wind structure imposed by the magnetic field.

The upper panels of Fig.~\ref{Fig:dens_sn} plot the density squared of
two simulation snapshots. They illustrate how below $r \approx R_{\rm
  A} \approx 2.7 R_\star$, the wind does indeed become trapped by the
closed field-line loops, whereby the material is pulled back by
gravity onto the star over a dynamical time-scale. But a key point
here is that, despite the very dynamical behaviour, the transient
suspension of material within such closed loops still results in a
wind region, in the vicinity of the magnetic equator, that
\textit{statistically} is overdense (Fig.~\ref{Fig:dens_sn},
lower-left panel). Further, as a result of the colliding wind material
at individual loop tops, this overdense region is also characterized
by very low velocities (Fig.~\ref{Fig:dens_sn}, lower-right panel), in
qualitative agreement with the narrowness of the observed $\Ha$
emission discussed in Sect.~\ref{spherical}.

The structures predicted by these simulations are physically distinct
from those predicted for rapidly rotating magnetic stars with $R_{\rm
  A} > R_{\rm K}$ \citep{Townsend05, Townsend07, udDoula08}, where
$R_{\rm K} = (v_{\rm rot}/v_{\rm crit})^{-2/3} R_\ast$ is the Kepler
co-rotation radius for critical rotation speed $v_{\rm crit}$. For
such stars, the centrifugal forces can support any trapped material
above $R_{\rm K}$, allowing then the magnetically confined wind to
accumulate material and form a \textit{centrifugal magnetosphere}
(CM). In contrast, the characteristic structure described above,
appropriate for slowly rotating massive stars with $R_{\rm K} > R_{\rm
  A} > R_\ast$, instead establishes the concept of a \textit{dynamical
  magnetosphere} (DM) \citep[see also][]{Petit11}.  

In hot coronae from the sun and some magnetically active cool stars,
there are analogous examples of regions of dynamical infall
\citep["coronal rain"; e.g.][]{Eibe99} or centrifugally supported
prominences \citep{Cameron03, Jardine05}, fed largely by the transient
eruptive propulsion of stellar flares.  By contrast, hot-star
magnetospheres are fed by the quasi-steady wind upflow driven by the
star's radiation, allowing for persistent Balmer emission that has
been monitored over multi-year time-scales spanning many rotation
periods.

Note that even rapid rotators will have a DM component at $r \la
R_{\rm K}$. But the $\Ha$ emission contribution from this part will be
insignificant because of the much higher densities at $R_{\rm K} < r <
R_{\rm A}$ \citep[see, e.g., Fig. 7 in][]{Townsend07}. These higher
densities stem from the much longer accumulation time-scale associated
with a CM \citep[typically months/years, see Appendix A
  in][]{Townsend05} than with a DM (typically hours, the dynamical
time-scale). But a star such as \hd, with $R_{\rm A} \approx 2.7
R_\star << R_{\rm K} \approx 55 R_\star$, only has a DM contributing
to the $\Ha$ emission. So whereas rapidly rotating magnetic B-stars
can indeed show substantial Balmer line emission, as observed in
e.g. $\sigma$ Ori E, the short accumulation time-scale of a DM
requires the relatively high mass-loss rate of an O-star to produce
observable $\Ha$ emission in these slowly rotating stars.

\subsection{Radiative transfer}
\label{RT}

To model the observed variation of Balmer emission, we compute
synthetic $\Ha$ flux profiles directly from the MHD simulations by
solving the formal integral of radiative transfer in a 3-D cylindrical
coordinate system ($p,\xi,z'$). This system is aligned toward the
observer by rotating the stellar system ($r,\theta,\phi$) by an angle
$\alpha$ about its y-axis, so that $\cos \alpha = \hat{z} \cdot
\hat{z}'$. The angle $\alpha$ thus defines the observer's viewing
angle with respect to the magnetic pole. For given $\beta$ and $i$,
we have
\beq
\cos \alpha = \sin \beta \cos \Phi \sin i + \cos \beta \cos i,
\eeq
which then readily gives the observer's viewing angle as function of
rotation phase $\Phi$. We note that even though the MHD models are
2-D, the radiative transfer must be performed in 3-D, as the axial
symmetry is broken for any observer with $|\cos \alpha| \ne 1$.

The transfer equation is solved only in the wind, with a pre-specified
photospheric profile $P_\nu$ as a lower boundary condition, taken from
NLTE model atmosphere calculations assuming negligible wind
contamination. While not truly self-consistent, this procedure has
been shown to be very accurate for $\Ha$ line profile calculations in
1-D smooth \citep{Puls96} as well as multi-dimensional clumped
\citep{Sundqvist11} O-star wind models without magnetic fields.

The monochromatic optical depth along a ray is 
\beq
\tau_\nu = \int \chi_\nu dz',   
\eeq
where $\chi_\nu$ is the frequency dependent opacity per unit
length. The opacities are calculated assuming an optically thin
continuum and occupation numbers for the $\Ha$ atomic levels $i$ given
in terms of the NLTE departure coefficients $b_{\rm i} = n_{\rm
  i}/n_\star$. Here $n_\star$ is the occupation number of level $i$ in
LTE with respect to the ground state of the next ionization state
\citep[e.g.,][]{Mihalas78}. The line profile is a Gaussian of Doppler
width $\Delta \nu_{\rm D}$, set by the local wind electron temperature
$T_{\rm e}$, and centered at zero co-moving frame frequency $x_{\rm
  cmf} = x_{\rm obs} - \hat{z}' \cdot \vec{v}/v_\infty$, where $x =
(\nu/\nu_0-1)c/v_\infty$.

The emergent intensity for a given ray then is  
\beq
I_\nu = P_\nu I_0 e^{-\tau_\nu^\infty} + \int_0^{\tau_\nu^\infty} S_\nu(\tau_\nu)
e^{-\tau_\nu} d \tau_\nu,
\eeq
where $I_0$ is the stellar photospheric continuum intensity, $S_\nu$
the NLTE line source function, and $\tau_\nu^\infty$ the optical depth
integrated over the complete ray. $I_0$ is taken from {\sc fastwind}
model atmospheres for rays that intersect the stellar core and set to
zero otherwise. $S_\nu$ is fixed by the $\Ha$ departure coefficients
and the local wind electron temperature. The emergent flux is then,
finally, obtained by integrating the emergent intensity over the
projected stellar disc.

\subsubsection{Electron temperatures and $\Ha$ occupation numbers}
\label{te_ni}

As described, the $\Ha$ synthesis problem requires estimates of
$T_{\rm e}$ and the hydrogen departure coefficients. But the energy
equation as treated in the MHD simulations described in
Sect.~\ref{MHD} yields only a rough approximation of the wind
temperature balance, with the local temperature artificially never
allowed to drop below a certain floor value (on the order of the
stellar effective temperature). In our $\Ha$ calculations, we
therefore estimate the wind temperature balance using the results of a
spherically symmetric {\sc fastwind} model, except in regions
shock-heated to $T_{\rm e} > 10^5 \rm K$, where we set the $\Ha$
opacity and source function to zero.

To consistently calculate NLTE departure coefficients for a full
multi-D MHD wind simulation is a daunting task, well beyond the scope
of the present paper. However, for now we take advantage of the fact
that hydrogen is almost fully ionized in typical O star winds. The
$\Ha$ line formation is then controlled by recombination, a thermal
process, and the participating atomic levels are therefore very close
to LTE with respect to ionized hydrogen. For \hd, {\sc fastwind}
calculations show deviations smaller than a factor of two, a typical
number for most `normal' O-star winds without strong magnetic fields
\citep{Puls96, Sundqvist11}. As a first approximation then, we here
take the simplest approach possible and assume LTE conditions, whereby
$b_{\rm i}=1$.

\section{$\Ha$ variability in a dynamical magnetosphere}
\label{results}
 
\begin{figure}
\includegraphics[width=4.5cm, angle=90]{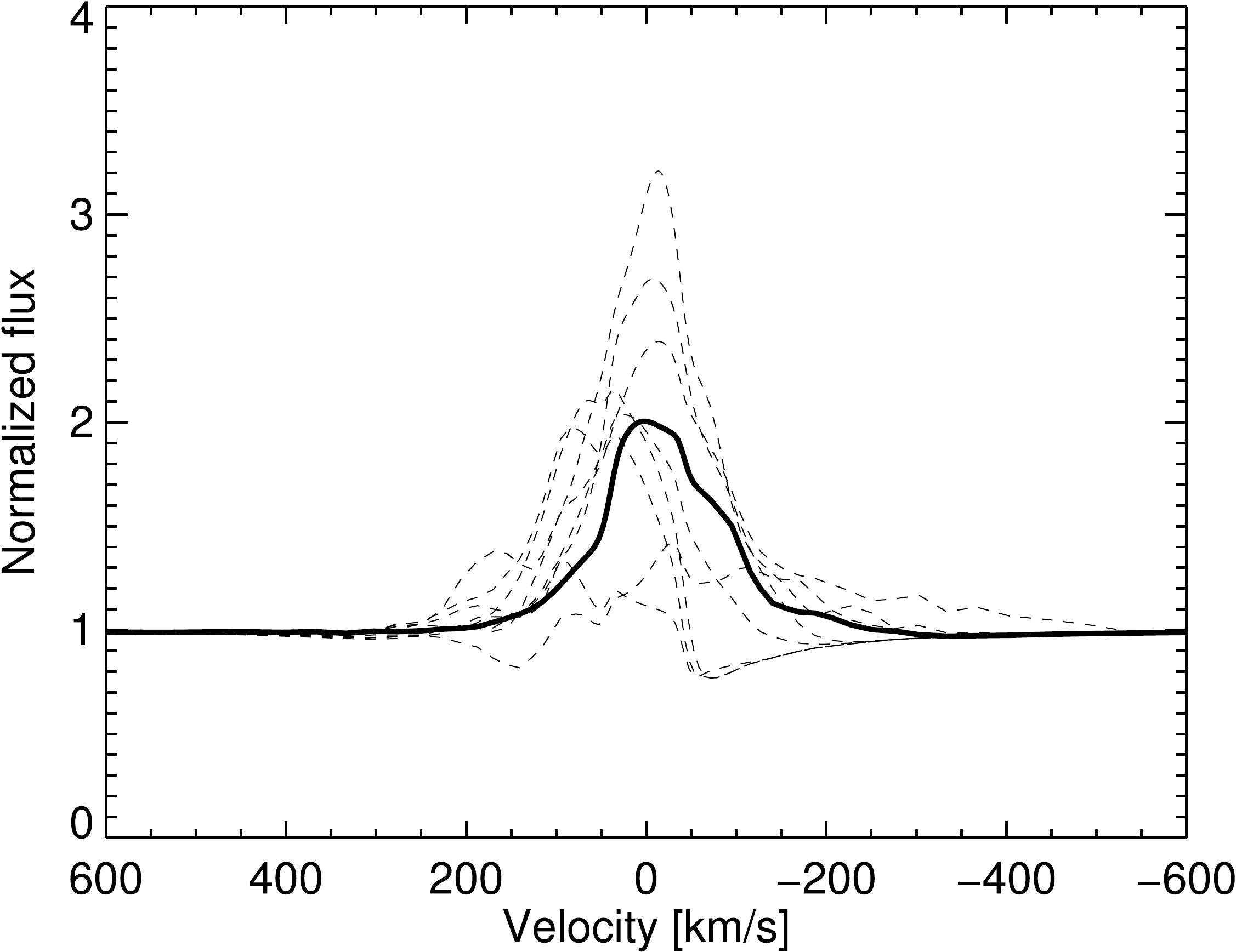}
\centering
\caption{Synthetic $\Ha$ profiles for 8 randomly selected snapshots
  during our simulation run (dashed lines), and a mean profile 
  calculated from them (solid line).}
\label{Fig:snaps}
\end{figure}

\begin{figure}
\begin{minipage}{8.0cm}
\resizebox{\hsize}{!}{\includegraphics[angle=90]{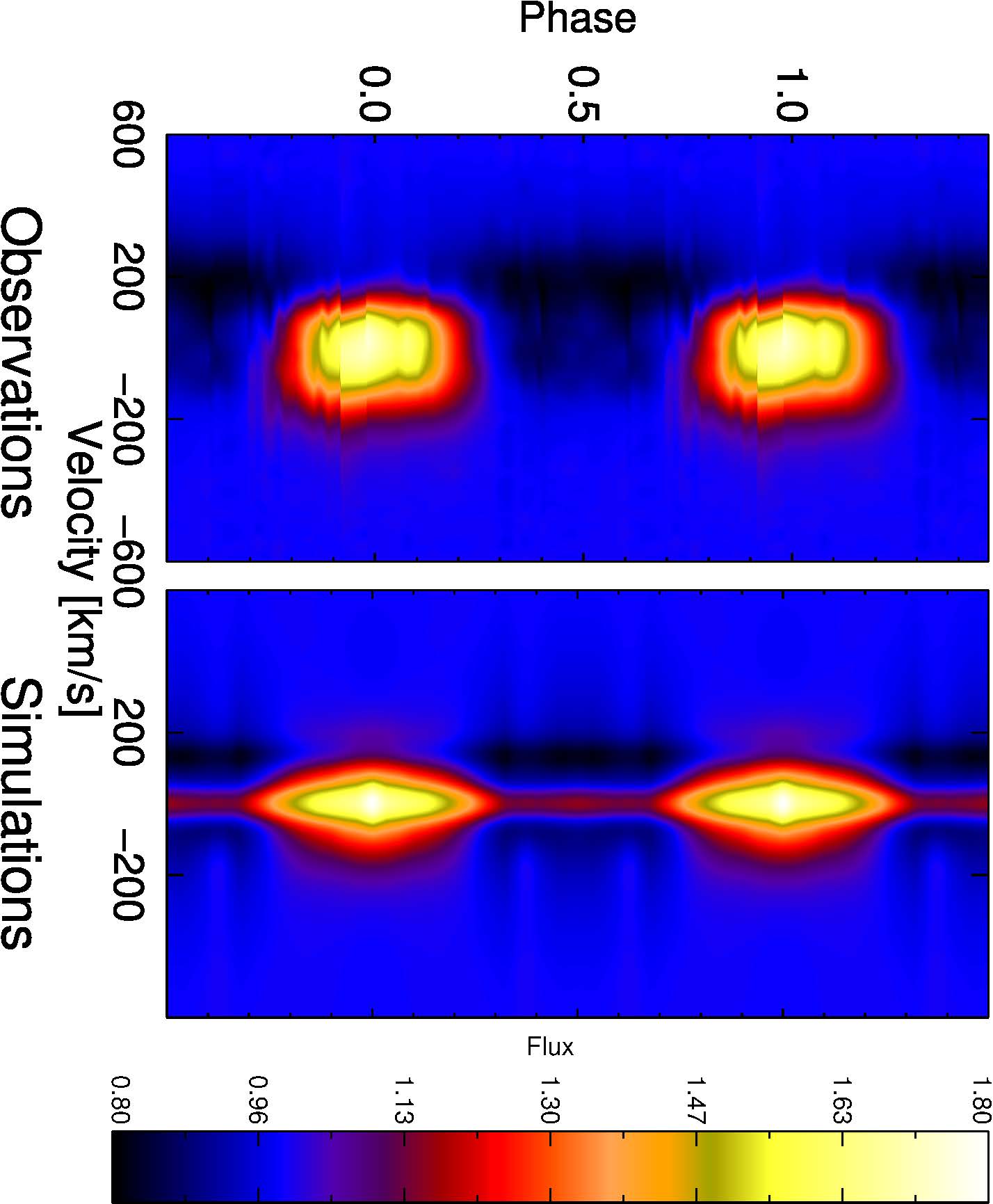}}
\end{minipage}

\begin{minipage}{8.0cm}
\resizebox{\hsize}{!}{\includegraphics[angle=90]{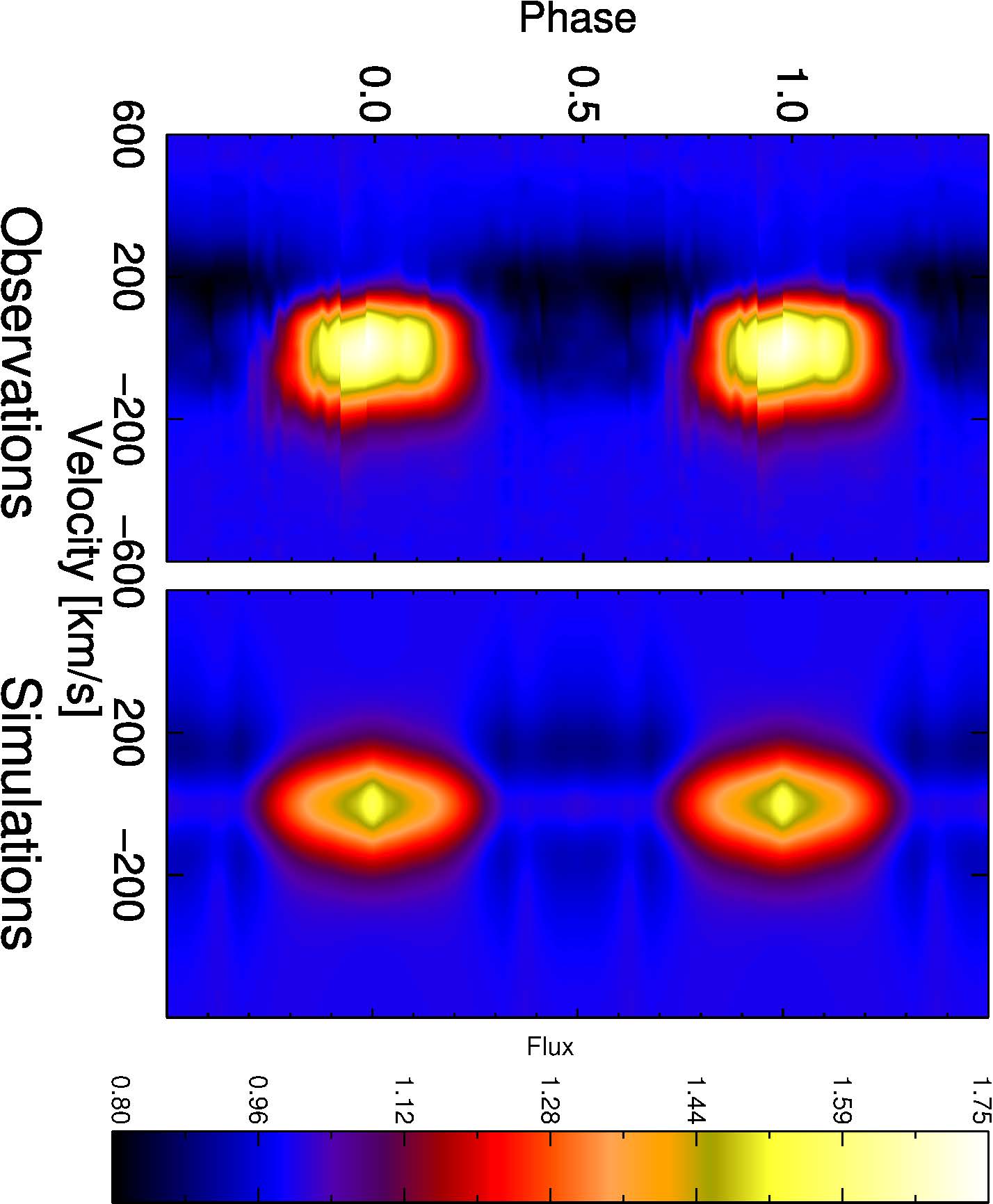}}
\caption{Observed \citep[ephemeris from][]{Howarth07} and synthetic
  $\Ha$ dynamic spectra, as functions of rotation phase and shown over
  two cycles. Synthetic spectra assume $\beta = i = 50 \degree$ and
  are calculated as described in text. The model spectra in the lower
  panel have further been convolved with an isotropic
  `macro-turbulence' of 100 km/s.}
\end{minipage}
\label{Fig:dyn}

\end{figure}

\begin{figure}
\resizebox{\hsize}{!}{\includegraphics[angle=90]{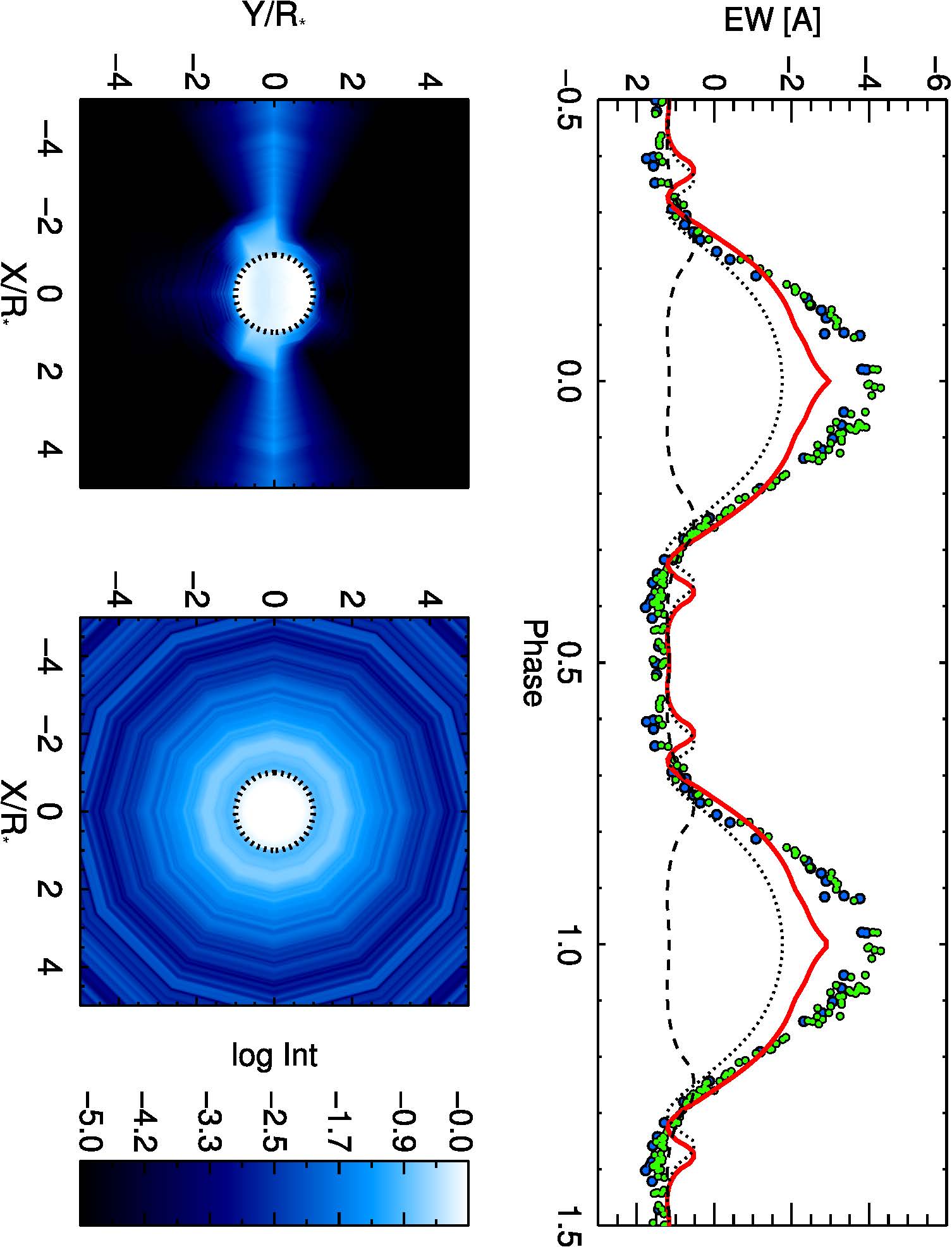}}
\caption{\textit{Upper panel:} Observed (filled circles) and simulated
  $\Ha$ equivalent widths as functions of rotation phase, for the
  three geometries $i=\beta=50 \degree$ (red, solid), $i=30 \degree,
  \, \beta=70 \degree$ (dotted), and $i=10 \degree, \, \beta=90
  \degree$ (dashed). The blue filled circles represent the higher
  quality data points used in Fig.~\ref{Fig:dyn}. \textit{Lower
    panels:} Projected surface area contours of the normalized
  emergent intensity at line center, for observers located along the
  magnetic equator (left) and pole (right). Cartesian coordinate $x'$
  on the abscissa and $y'$ on the ordinate.}
\label{Fig:sb}
\end{figure}

We calculate an $\Ha$ line profile for more than 100 snapshots of the
2-D MHD simulation of \hd. Fig.~\ref{Fig:snaps} shows that such
profiles are highly variable, mimicking the wind's dynamical
nature. To obtain a mean profile at each phase, we average over such
profiles for $\sim$100 randomly selected snapshots; this is intended
to be a simple proxy for the real 3-D nature of the wind dynamics,
effectively using the complex and non-linear variations in time in our
2-D simulation to mimic the expected variations in azimuth in a full
3-D model.\footnote{For this type of line transfer, which often is
  optically thick, such post-averaging of the line profiles computed
  for many time snapshots is more realistic than the simple
  pre-processed, time-averaged wind density used by \citet{Wade11a} to
  model polarized electron scattering, which is more nearly optically
  thin.}  While approximate, this seems a reasonable first approach to
account for the limited lateral coherence or synchronization of an
actual 3-D magnetized wind. Also, each phase is constructed by further
averaging the two phases having equal $\Phi$'s when reversing the
magnetic poles, to ensure the expected long-term north/south symmetry
of our simulation. The procedures described above effectively smooth
out most of the short-time variability of our simulation, in agreement
with the observations \citep{Howarth07}.

Fig.~\ref{Fig:dyn} compares observed and synthetic time-averaged
dynamic $\Ha$ spectra, plotted as functions of rotational phase
assuming $\beta=i=50 \, \degree$. This is consistent with the $\beta +
i = 95 \pm 10 \degree$ derived by \citet{Wade11a}, but differs
slightly from the $i = 30 \degree$ adopted there (see further
below). The observed general trends, with peak flux at phase 0 and an
extended minimum around phase 0.5, are both well reproduced. The flux
variations are caused by differences in the projected surface area of
overdense $\Ha$ emitting material as the observer changes viewing
angle when the star rotates.  Fig.~\ref{Fig:sb} demonstrates this by
plotting the $\Ha$ emergent intensity (surface brightness) at line
center for observers located along the axes of magnetic pole and
equator. The figure clearly shows how the flux, which is just the
integral of the intensity over this projected area, is much higher for
the observer along the polar axis.

The large observed $\Ha$ variability puts rather tight constraints on
the system's geometry. The equivalent width curves in
Fig.~\ref{Fig:sb} directly refute very low values of $i$, but also
show that the $i=30 \degree$, $\beta = 70 \degree$ assigned as a
tentative reference geometry by \citet{Wade11a} results in somewhat
weaker variation than the $\beta = i = 50 \degree$ adopted here. This
is simply because, for a given sum $i + \beta = 100 \degree$, an
observer at $i = 30 \degree$ never looks closer to the magnetic pole
than $\alpha \approx 40 \degree$, whereas for an observer at $i = 50
\degree$, $\alpha$ spans the entire range from pole to equator, and
back again, in one rotation period, thus resulting in larger flux
variations.
 
There are discrepancies, of course. Whilst the observed and simulated
equivalent width curves qualitatively agree well, the simulated
variation is quantitatively somewhat too low. These deviations could
however be remedied if the DM were more concentrated, which would
result in a larger surface brightness difference between pole and
equator (Fig.~\ref{Fig:sb}). Such stronger wind confinement could
occur from either a lower mass-loss rate or a stronger magnetic field,
where the former choice seems more likely (due to clumping,
Sect.~\ref{MHD}), since the $i = 50 \degree$ adopted here actually
would result in a slightly reduced magnetic field strength as compared
to that derived by \citet{Wade11a}, who adopted $i = 30
\degree$. Another possibility is of course that these slight
discrepancies simply are related to insufficient assumptions for the
wind electron temperature structure and/or the hydrogen occupation
numbers (Sect.~\ref{te_ni}).

In addition, the velocity dispersion in the models is too low,
predicting narrower and sharper peaked profiles than observed
(Fig.~\ref{Fig:dyn}, upper panel). To illustrate this further, the
lower panel of Fig.~\ref{Fig:dyn} displays the same line profiles as
the upper panel, but now with the model profiles convolved with an
isotropic Gaussian `macro-turbulence' of 100\,km/s. Inspection shows
that the new profile shapes are significantly improved. We suspect
that this required macro-turbulence is an artefact of missing wind
dynamics, since, generally, the added degree of freedom in a 3-D
simulation should result in larger velocity dispersion than in the 2-D
models employed here. Though very challenging, 3-D MHD models are
currently being developed by one of us (A.~ud-Doula), and will be
reported in a future paper. Such 3-D simulations will then also
quantitatively test our simple approach here of averaging radiative
transfer results for 2-D simulation snapshots to approximate the real
3-D dynamical wind structure.

Finally, notwithstanding the foregoing comments, there is surprisingly
good overall agreement between the models and observations, which
strongly supports that the DM model captures the key physics
responsible for the $\Ha$ variability.

    
\section{Discussion and conclusions}
\label{discussion}

We have demonstrated that radiation magneto-hydrodynamical simulations
of a confined wind, together with detailed radiative transfer
modelling, reproduce well the distinct periodic \Ha~emission observed
in the magnetic O-star \hd. We interpret this within the context of a
\textit{dynamical magnetosphere} (DM), wherein the rotationally
modulated spectral variations are results of a statistically
overdense, low velocity wind region around the magnetic equator.

While applied here only to \hd, the DM model may also well describe
optical Balmer line variability in other magnetic O stars with $R_{\rm
  K} > R_{\rm A}$, such as $\theta^1$ Ori C and HD\,148937
\citep{Wade06, Wade11b}. Indeed, the narrow \Ha~emission observed in
HD\,148937 suggests a line formation scenario in a DM, with the only
significant difference to \hd~then being the stellar and/or magnetic
geometry \citep{Wade11b}, resulting in much smaller spectral
variations for the former star. In future work, we intend to develop
further this DM model and apply it to a broader sample of magnetic
massive stars that show variable Balmer emission and are characterized
by $R_{\rm K} > R_{\rm A}$.

\section*{Acknowledgments}

G.A.W. acknowledges support from the Natural Science and Engineering
Research Council of Canada. This work was supported in part by NASA 
ATP grant MANX11AC40G.


\bibliography{sundqvist}

\end{document}